\documentclass[a4paper,12pt]{article}
\usepackage{mathtext,amssymb,graphicx}
\usepackage{epsfig}
\usepackage{caption}
\usepackage[T2A]{fontenc}
\usepackage[utf8]{inputenc}
\newcommand{\be}{\begin{equation}}
\newcommand{\ee}{\end{equation}}
\newcommand{\nc}{\newcommand}
\nc{\bi}{\bibitem}
\newcommand{\ba}{\begin{array}}
\newcommand{\ea}{\end{array}}

\def\f{\frac}

\def\a{\alpha}
\newcommand{\beq}{\begin{equation}}
\newcommand{\eeq}{\end{equation}}

\newcommand{\myref}[1]{~{(\ref{#1})}}

\newcommand{\myfigref}[1]{~{Fig.~(\ref{#1})}}

\makeatletter
\newcommand\figcaption{\def\@captype{figure}\caption}
\makeatother

\begin{document}

\begin{titlepage}

\hfill\parbox{40mm}
{\begin{flushleft}  ITEP-TH-10/09\\
FTPI-MINN-09/12
\end{flushleft}}

\vspace{10mm}

\centerline{\large \bf Correlator of Wilson and t'Hooft Loops}
\centerline{\large \bf at Strong Coupling in $\mathcal{N}=4$ SYM Theory}

\vspace{17mm}

\centerline{\bf A.~Gorsky$^{1,2}$, A.~Monin$^{4}$,
A.V.~Zayakin$^{1,3}$}
\begin{center}
{\it $^1$ Institute of Theoretical and Experimental Physics, Moscow, Russia, \\
$^2$ FTPI, University of Minnesota.} \\
$^3$ {\it Department f\"ur Physik der Ludwig-Maximillians-Universit\"at M\"unchen und\\
Maier-Leibniz-Labor, Am Coulombwall 1, 85748 Garching, Germany\\
$^4$ School of Physics and Astronomy, University of Minnesota}
\vspace{1cm}
\end{center}

\centerline{\bf Abstract}
We calculate the correlator of a 't Hooft and a Wilson coplanar circular concentric loops
at strong coupling in $N=4$ SYM theory.  In this limit the problem reduces to the
determination of the
composite minimal surface
in the curved space with the proper boundary conditions.
The minimal admissible ratio of radii for such a configuration is found to be  $\mathrm{e}^{-1/2}\approx 0.606$ at zero
temperature  and  the dependence of the minimal admissible radii ratio on
temperature is derived. At low temperatures the minimal admissible ratio of 't Hooft and Wilson loops radii
remains close to $0.6$, whereas at high temperatures $T$ it becomes equal to $\frac{1}{\pi T}$.
We find that at any temperature there exists a phase transition point: beneath some specific
value of 't Hooft loop radius the dual counterpart of Wilson--'t Hooft correlator is
organized as two disconnected surfaces in AdS, whereas for 't Hooft loop radius above it,
there exists a connected configuration with a junction of monopole, charge and dyon surfaces.
We suggest a generalization of the entanglement entropy for charged boundaries and make some
comments on its calculation at strong coupling.


\end{titlepage}

\section{Introduction}

There is growing interest to the possible role of the magnetic degrees of freedom in the
strongly coupled gauge theory at
nonzero temperature. It was argued that monopole-like degrees of freedom can be relevant
for the description
of the quark-gluon plasma and even a possibility of the ``selfdual'' plasma  has been discussed.
The current status of this issue and the proper references can be found in  \cite{shuryak,cz}. Hence
it is interesting to investigate the interaction between the magnetic and electric degrees of freedom
in details. There are in principle two different mechanisms of interaction; one involves an
exchange by a particle-like mode while the second way can be better thought of as the stringy one.
We shall focus in this paper on the stringy mechanism of interaction and shall discuss it in the
dual picture relevant to the strong coupling regime.

In the dual stringy setup the presence of the $(p,q)$ string junctions known in the $IIB$ model is of prime importance. The configuration involving junction  was successfully applied to derivation of the quark-monopole
interaction \cite{minahan} when the dual description  of the parallel straight electric and magnetic lines
at the boundary  was considered. Another application of the junction concerns the nonperturbative
decay of the monopole
in the  electric field and electrically charged particle in the  magnetic field.
Such processes were suggested in \cite{gss} and analyzed in details in \cite{mz}. In all cases
the configuration involves some ``virtual'' dyons which are coupled with external particles at
the junction manifold where the magnetic and electric charges are conserved.

In this paper we shall consider the correlator of a Wilson and a t'Hooft loops which was suggested
in \cite{gz} as a good probe for the investigation of the magnetic properties of the plasma.
In the dual picture the Wilson loop in the strong coupling regime is calculated in terms of the
minimal surface of the fundamental string with Wilson loop boundary \cite{maldacena} while the
t'Hooft loop is the boundary of the D-string worldsheet. To simplify the consideration we shall discuss
the circular concentric  magnetic and electric loops and
shall search for the connected surface with the proper boundary conditions. It is clear that due to the charge conservation the dyonic string
worldsheet must be involved and the minimal surface has to be composite.  We shall find such
composite minimal surface involving the worldsheets of F1, D1 and (F1,D1) strings in some  interval of the
radii ratio.

It is typical that the stringy minimal surfaces involving several
boundaries undergo a kind of phase transitions familiar from soap
films. In the dual description such critical behavior was first
discussed in \cite{gross} for the parallel Wilson loops in the
boundary theory. Later similar transitions have been found for the
coplanar circular Wilson loops \cite{olesen}.  In our analysis of
the composite minimal surface we shall find the similar phase
transition. Above some critical ratio of electric and magnetic radii
the connected surface is absent,  that is, the correlator can be
saturated only by the exchange of some particular supergravity
modes. We find the dependence of the minimal admissible radii ratio
on temperature. At low temperatures the minimal admissible ratio of
't Hooft and Wilson loops remains close to $0.6$, whereas at high
temperatures $T$ it becomes equal to $\frac{1}{\pi TL}$, where $L$
is AdS radius.

Another related issue we shall discuss concerns the entanglement entropy  which effectively counts the number of
degrees of freedom stored in some particular region. The holographic calculation of the
entanglement is very similar
to the calculation of the Wilson loop correlators since it is nothing but the calculation of the
minimal surface in the bulk with the proper boundary \cite{ryu}. The entanglement entropy manifests
some critical behavior as the function of the size of the region which suggests that it can serve as a kind
of order parameter for the deconfinement phase transition \cite{klebanov}. Another important feature
of the entanglement entropy is the property of the strong subadditivity for the region with the
multiple boundaries \cite{strong}. It can be proved holographically \cite{Headrick:2007km} and corresponds
to the inequalities for the Wilson loops in the gauge theories \cite{bachas}.

In our analysis we shall meet a new situation when the boundaries  carry electric, magnetic or dyonic charges.
The calculation of the entanglement entropy in the annulus geometry has been considered in \cite{hirata2}
when the smooth minimal surface connects two boundaries. In our case this is impossible because of the
charge conservation that is why we shall suggest the generalization of the entanglement entropy
for the regions with the charged boundaries. Our evident recipe for the dual calculation of the charged entanglement
entropy involves the composite minimal surface we have found. We shall briefly discuss some generic
properties of the charged entanglement entropy,  like the strong subadditivity.

The paper is organized as follows. In Section 2 we shall find the composite minimal surface at zero
temperature and discuss its properties. In Section 3 we consider the dependence of the composite
minimal surface on temperature, which is introduced by background modification. Section 4
is devoted to our suggestion for the generalized entanglement entropy and we
investigate some of its properties.
We summarize our findings and list the open questions in the Conclusion.

\section{Zero-temperature configuration}
In this section we shall investigate the correlator
of electric and magnetic loops at zero temperature. Since
we perform a strong coupling calculation, the problem is reduced
to calculating the minimal surface in $AdS_5$
background~\cite{maldacena}
with an electrically and a magnetically charged boundary.
Contrary to the correlator of the similar loops we have to add
a ``virtual'' dyonic surface to provide  charge conservation.
Another extra element is the account of the equilibrium condition
for the tensions at the junction line.

\begin{figure}[t]
\begin{center}
\includegraphics[height = 4.1cm, width=10cm]{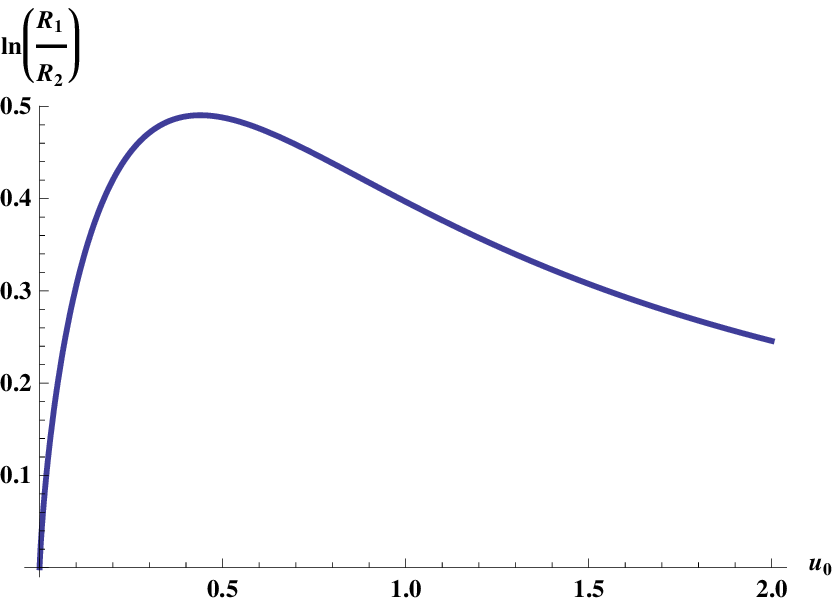}
\caption{Right-hand-side of equation\myref{intersect}, $t=100$\label{rhs}}
\end{center}
\end{figure}

Our metric is
\beq
ds^2=\frac{dz^2+dr^2}{z^2}.
\eeq
and we start with the action for an axially-symmetric world-sheet in terms of $r(z)$ as dynamical variable.
Here and below all dimensionful quantities are measured in units of AdS radius $L$.
For a string with tension $T$, where
\beq T=\left\{
\begin{array}{ll}
1, &\mbox{charge (Wilson)}\\
t, &\mbox{monopole ('t Hooft)}\\
\sqrt{1+t^2},&\mbox{dyon},
\end{array}\right.
\eeq
the action is
\beq
S=T\int \frac{r dz }{z^2}\sqrt{1+r^{\prime 2}}.
\eeq
This action does not make the conformal symmetry of the problem manifest. To make it explicit, we make a substitution $(z,r(z))\to (u, \tau(u))$, where
\beq
\begin{array}{l}
z=\frac{e^\tau}{\sqrt{1+u^2}},\\
r=\frac{e^\tau u}{\sqrt{1+u^2}}.
\end{array}
\eeq
In these variables
\beq
ds^2=\frac{du^2}{1+u^2}+(1+u^2)d\tau^2,
\eeq
and
\beq
S=\int\frac{udu}{\sqrt{1+u^2}}\sqrt{(1+u^2)^2\tau^{\prime 2} +1}.
\eeq
Variables $u,\tau$ are ``action---angle'' variables for conformal symmetry, which now manifests itself in $\tau$ shift invariance. Correspondingly, it induces integral of motion $c=\frac{\partial L}{\partial \tau^\prime}$, which allows us to express
\beq
\tau^\prime=\pm \frac{c}{(1+u^2)\sqrt{u^4+u^2-c^2}}.
\eeq
This immediately brings an expression for a normalized tangential vector $n^\mu\equiv (n^u,n^\tau)$
\beq
n^\mu=\left(\frac{\sqrt{u^4+u^2-c^2}}{u},
\frac{c}{u(1+u^2)}\right).
\eeq
Solution to equations of motion is given in terms of the elliptic incomplete function $\Pi(a;z|b)$
\beq
\begin{array}{rcl}
f(\tau,c)&=&\int \frac{c du}{(1+u^2)\sqrt{u^4+u^2-c^2}}\equiv\\
&\equiv&\frac{\sqrt{2} c \Pi \left(\frac{1}{2} \left(\sqrt{4 c^2+1}+1\right);\,\,i \sinh ^{-1}\left(\frac{\sqrt{2} u}{\sqrt{\sqrt{4
   c^2+1}+1}}\right)|\frac{\sqrt{4 c^2+1}+1}{1-\sqrt{4 c^2+1}}\right)}{\sqrt{\sqrt{4 c^2+1}-1}}.
\end{array}
\eeq
The configuration we are interested in is described by three curves $\tau(u;c)$. One is a dyon and is simply given as $c=0$, that is
\beq
\tau=const,
\eeq
making a sphere $z^2+r^2=const^2$ in terms of $(z,r)$. The other two lines are a charge $\tau_2(u;c_2)$ and a monopole $\tau_1(u;c_2)$, whose parameters are constrained by boundary conditions of the Wilson and 't Hooft lines, making circles of radius $R_2,R_1$ respectively, where $R_2>R_1$:
\beq
\tau_1=\log R_1 -F(u;c_1),
\eeq
and
\beq
\tau_2=\log R_2 +F(u;c_2),
\eeq
and $F(u;c)\equiv f(
u,c)-f(\infty,c)$.
We have three more constraints upon the configuration: force equilibrium conditions and intersection condition. From the equilibrium of forces we obtain
\beq
c_2=t c_1,
\eeq
and
\beq
\frac{\sqrt{u^4+u^2-c_2^2}}{u}+
t\frac{\sqrt{u^4+u^2-c_1^2}}{u}
=\sqrt{1+t^2}\sqrt{1+u^2},
\eeq
which yields
\beq
c_1=\pm u\sqrt{\frac{1+u^2}{1+t^2}},
\eeq
and
\beq
c_2=\pm t u\sqrt{\frac{1+u^2}{1+t^2}}.
\eeq
The third condition
\beq\label{intersect}
\log \frac{R_2}{R_1}=-F(u;c_2)-F(u;c_1)
\eeq
can be easily solved numerically. Its right-hand-side is depicted in \myfigref{rhs}. We can see that, for all couplings, it has two regular solutions, but only for $ 0<\log\frac{R_2}{R_1}<\frac{1}{2}$. Thus the configuration we are looking for, exists only for a limited range of $\frac{R_2}{R_1}$.

Regularized (UV finite) action on equations of motion is given by
\beq
\sigma(u;c)=\int \left(\frac{u^2}{\sqrt{u^4+u^2-c^2}}-1\right)du
\eeq
Thus
\beq
S=\sqrt{1+t^2}\left(\sqrt{1+u_0^2}-u_0-1\right)+
\left(\sigma(\infty;c_2)-\sigma(u_0;c_2))+
t(\sigma(\infty;c_1)-\sigma(u_0;c_1)\right)
\eeq

\begin{figure}[t]
\begin{center}
\includegraphics[height = 6cm, width=7cm]{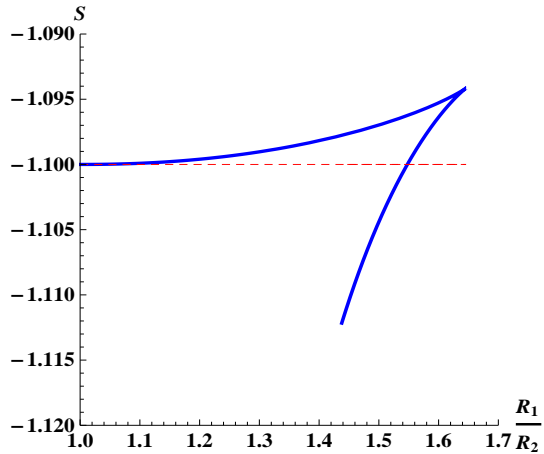}
\caption{\label{action} Action in the units of $\f {R ^2 _{AdS}} {\a'}$ vs. ratio $R_1/R_2$ for the connected configuration (solid line), and for the disconnected configuration (dashed line); $t=1/10$}
\end{center}
\end{figure}

\begin{figure}[t]
\begin{center}
\includegraphics[height = 4.1cm, width=10cm]{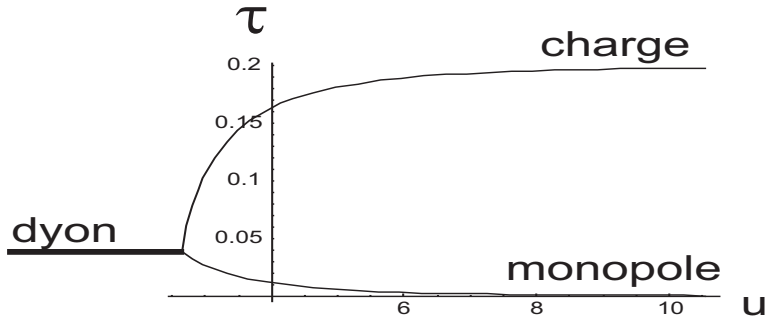}
\caption{\label{chargemonopoledyon} Configuration of charge, monopole and dyon lines, $t=3$.}
\end{center}
\end{figure}

Comparing the action on the two possible solutions, we can see in\myfigref{action} that the action on solution 1 decreases when $\log\frac{R_2}{R_1}\to 0$, whereas action on solution 2 decreases when $\log\frac{R_2}{R_1}\to 1/2$.
The corresponding configurations in $u, \tau$ space look like shown in \myfigref{chargemonopoledyon}.

\section{Finite-temperature analysis}

\begin{figure}[t]
\begin{center}
\includegraphics[height = 5.5cm, width=7cm]{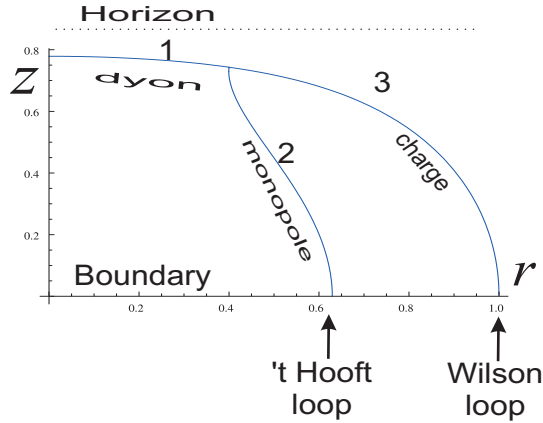}
\caption{\label{config} Connected configuration for Wilson--'t Hooft correlators at finite temperature and $g\to \infty$.}
\end{center}
\end{figure}

\begin{figure}[t]
\begin{center}
\includegraphics[height = 5.5cm, width=8cm]{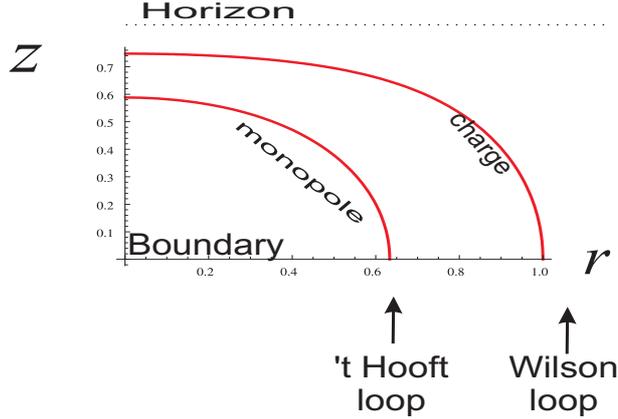}
\caption{\label{config1} Disconnected configuration of Wilson--'t Hooft correlators at finite temperature and $g\to \infty$.}
\end{center}
\end{figure}

 In this Section we generalize the zero-temperature analysis to finite-temperature case
to study the transition phenomena in this system, which is the main objective of this paper.
Let us take the thermal metric suggested in \cite{witten} which reads as
\beq
ds^2=\frac{R^2}{z^2}\left(-h(z)dt^2+dx_i^2 \right)
+\frac{R^2}{h(z)}\frac{dz^2}{z^2}
\eeq
where
\beq
h(z)=\left\{\begin{array}{l}
1-\frac{z^4}{z_0^4},\,z_0>2/3\\
z^2+1-\mu z^4,\,z_0<2/3
\end{array}\right.
\eeq
and $z_0=\frac{1}{\pi T}$, where  $T$ is the temperature. This metric corresponds either to compactified AdS geometry ($z_0>2/3$) or to a black hole in AdS ($z_0<2/3$). The type of the configuration we want to study remains the same as in the previous Section: Wilson and t'Hooft loops, connected with a dyon ``cap''.  For simplicity we study configurations with $g\to \infty$, which is consistent with duality hypothesis. Therefore, tension of monopole surface goes to zero and it is attached to the dyon-charge surface along a normal vector; the dyon-charge surface itself has no sharp bend.  The configuration we consider now is shown in\myfigref{config}.
Alternatively, a purely disconnected configuration may exist, consisting only of dyon and monopole cups, shown in\myfigref{config1}
The action on any of the surfaces, given by $r(z)$ dependence, is then
\beq
S=T_{p,q}\int dz \frac{r}{z^2}\sqrt{\frac{1}{h(z)}
+r^{\prime 2}}.
\eeq
As we will see below, the two configurations will be concurring for being the leading one. We shall establish criteria for existence and leadership of these configurations below. We note that in the limit $g\to \infty$ the only difference in action comes from the piece of the configuration, denoted (2) in\myfigref{config}, that is, the ``monopole'' surface, bounded by 't Hooft loop.

We solve the equations of motion numerically and construct a family of solutions $\{z_1(r;R_1,R_2,z_0,g),\,z_2(r;R_1,R_2,z_0,g),\,
z_3(r;R_1,R_2,z_0,g)\}$  subject to boundary conditions:
\beq\left\{
\begin{array}{ll}
z_1^\prime(0)=0 &\mbox{(regularity at origin)}\\
z_2(R_2)=0 &\mbox{('t Hooft line is a boundary)}\\
z_3(R_1)=0 &\mbox{(Wilson line is a boundary)}\\
z_1(\bar{r})=z_2(\bar{r})=z_3(\bar{r}) &\mbox{(junction present)}\\
T_{1,1}\tau_1^\mu(\bar{r})+T_{0,1}\tau_2^\mu(\bar{r})+
T_{1,0}\tau_3^\mu(\bar{r})=0&\mbox{(junction is in equilibrium)}
\end{array}\right.
\eeq
where $\bar{r}$ is the junction coordinate $r$, $\tau_i^\mu$ are tangential vectors to each of the respective curves $z_i(r)$, $\tau^\mu_i=\frac{1}{\sqrt{g_{zz}z_i^{\prime 2}+g_{rr}}}\left(1,z_i^\prime\right)$, indices $i=1,2,3$ refer to the parts of the configuration shown in\myfigref{config} . The use  of strong-coupling limit makes the equilibrium condition easy to implement numerically: the requirement is now simply $g_{\mu\nu}\tau_1^\mu\tau_2^\nu=0$.

Although any range of $(R_1,R_2)$ can be studied by our method, for definiteness we keep everywhere below the Wilson radius very large $R_1=1$, which is reasonable, since we want to see something interesting around phase transition, taking place at a comparable scale of  $z_0=2/3$. At zero temperatures as we have seen in the previous Section, all the dynamics of the system depends only on the ratio of $R_2/R_1$. At finite temperature it is already not the case, for conformal symmetry has been broken. Thus the action is now a function of two independent variables $R_1$ and $R_2$.  However, it seems justified to study the effects around $R_1=1$ for the following reason: a very small configuration will feel non-trivially only very high temperatures; similarly, a very large configuration will feel non-trivially only high temperatures; if anything interesting happens around Hawking--Page phase transition, it must be probed with $R_1\sim 1$.

\begin{figure}[t]
\begin{center}
\includegraphics[height = 8.1cm, width=8cm]{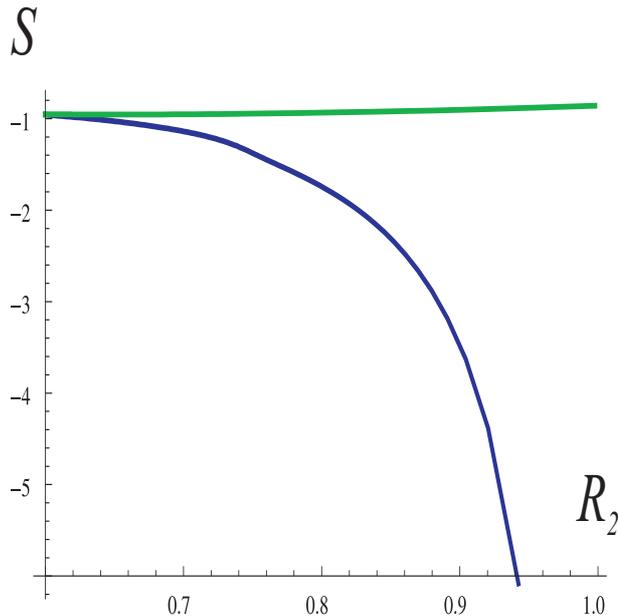}
\caption{\label{twobranches} Two branches of the action on the connected configuration, stable and unstable one.}
\end{center}
\end{figure}

\begin{figure}[t]
\begin{center}
\includegraphics[height = 5.1cm, width=8cm]{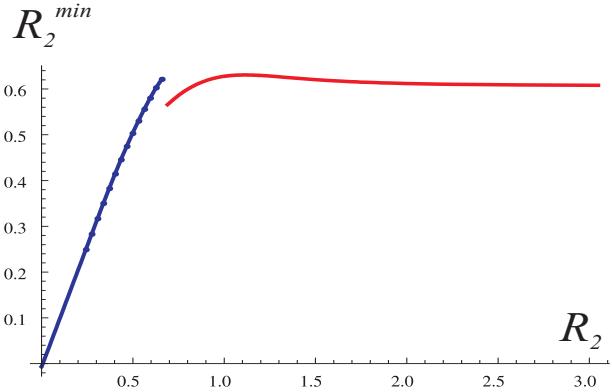}
\caption{\label{r2min} Minimal admissible 't Hooft loop radius $R_2^{min}(z_0)$ with Wilson loop being set $R_1=1$.}
\end{center}
\end{figure}

Upon solving the equations we observe that there are two branches of $S(R_2)$, a stable and an unstable one, shown in\myfigref{twobranches}. We note that our numerical recipe for finite renormalized action was to employ a constant cutoff.
The unstable branch is of no interest to us, because it never crosses the stable one.
The range of $R_2$ is here $(R_2^{min}(z_0),R_1)$. At zero temperature we have seen that $R_2^{min}=R_1e^{-\frac{1}{2}}\approx 0.6065 R_1$. At finite temperature the minimal admissible $R_2$ will be some function $R_2^{min}(z_0)$. It is shown in\myfigref{r2min}.
Smooth asymptote at $z_0\to \infty$ nicely reproduces the analytic zero-temperature result. In the black-hole temperature range the minimal radius is at a high accuracy $R_2^{min}(z_0)=z_0$.

\begin{figure}[t]
\begin{center}
\includegraphics[height = 4.1cm, width=8cm]{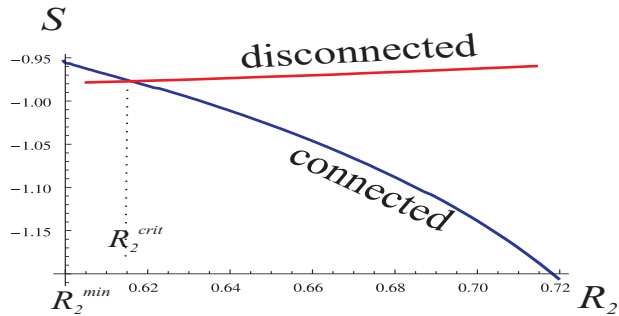}
\caption{\label{phasetransition} Below $R_2^{crit}$ disconnected piece dominates, above it --- a connected configuration has a lower action.}
\end{center}
\end{figure}

\begin{figure}[t]
\begin{center}
\includegraphics[height = 5.1cm, width=8cm]{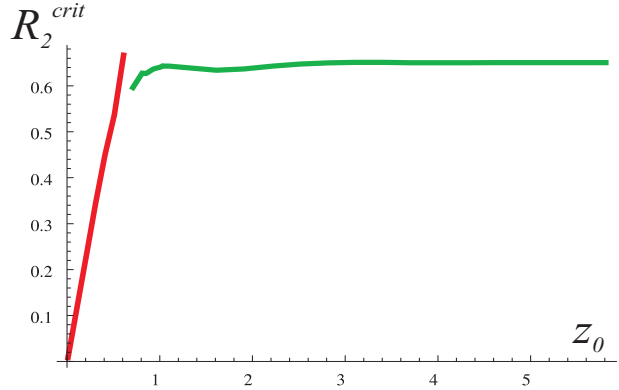}
\caption{\label{r2crit} Dependence of critical 't Hooft radius
$R_2^{crit}$ on inverse temperature parameter $z_0$.}
\end{center}
\end{figure}

Action evaluated on this branch should be compared to the action on the stable branch of the connected solution, which we consider in \myref{phasetransition}.
One can observe a critical point $R_2^{crit}$, distinct from $R_2^{min}$. The whole picture resembles
here the one found in the AdS/CFT computation of entanglement entropy \cite{klebanov}:
upper unstable connected branch is fully irrelevant, lower connected branch intersects with the disconnected solution at some critical scale $l_{crit}$, whereas a maximum admissible scale $l_{max}$ for a connected solution exists and is distinct from $l_{crit}$. We show the dynamics of $R_2^{crit}(z_0)$ in\myfigref{r2crit}.
Our analysis shows that this critical value is always close to the minimal admissible value, however, it is always slightly above it. Thus at all temperatures three exists a phase transition in the 't Hooft radius $R_2$.

\section{Generalized Entanglement entropy}

In this section we shall discuss the possible generalization of the notion of the
entanglement entropy for the regions with the charged boundaries suggested
by the analysis of the previous sections. This notion has been introduced long time ago
(see \cite{cardy} for the review and references) but attracted a lot of attention recently because
of its effective derivation in the holographic picture \cite{ryu}. Roughly speaking if we have a set of regions
divided by boundaries than the entanglement entropy is defined as the entropy seen by an observer in a region
who does not communicate with the other regions. In the simplest case  one has two regions A and B
and introduce the vacuum density matrix $\rho_0=|0><0|$.  Then the reduced density matrix
\beq
\rho_A= Tr_{B} \rho_0
\eeq
defines the entanglement entropy
\beq
S_A=- Tr_{A} \rho_{A} log\rho_a.
\eeq
The entanglement entropy is generically
UV divergent but the UV divergent part of the entropy does not depend on the size of the
region $L$ hence the finite  $L$-dependent contribution to the entanglement entropy
can be, for instance, safely defined as the difference of the entropies at two different $L_1$ and $L_2$.

The multicomponent regions has been investigated as well and the following generalization has been
suggested, inspired by the one-dimensional  case \cite{hubeny}
\beq
S(X_1 \cup X_2 \dots X_p) = \sum_{i,j} S_(|a_i -b_j|) - \sum_{i<j} S_(|a_i -a_j|)- \sum_{i<j} S_(|b_i -b_j|),
\eeq
where S is the entropy of the single component and $a_i$ and $b_i$ are the right and left boundaries of the
$i$-th component. An important question concerns the property of the strong subadditivity
\beq
S_A + S_B \ge S_{A\bigcup B} +S_{A\bigcap B},
\eeq
which has been proven in holographic picture in  \cite{hirata2}. Another interesting feature of the system to study
is the extensive mutual information \cite{casini}
\beq
I(A,B\cup C)= I(A,C) +I(A,C),
\eeq
where
\beq
I(A,B)= S(A) +S(B) -S(A\cup B).
\eeq
It was argued in \cite{casini} that the extensivity does not generically hold  which is triggered
by nonvanishing tripartite information function
\beq
I(A,B,C)= I(A,B) +I(A,C) -I(A,B\cup C).
\eeq

The holographic calculation of the entanglement entropy is practically identical to the Wilson loop calculation
hence our mixed correlators suggest the natural generalization of the entanglement entropy when the charges $(p_i,q_i)$
are attributed to each boundary. That is, the entropy function for each interval takes values
in $SL(2,Z)\otimes SL(2,Z)$ lattice  and has the following structure
\beq
S_i= S_{(p_i,q_i)}^{(p_{i+1}, q_{i+1})}
\eeq
for the $i$-th interval.  In the conformal case the calculation of the generalized entropy corresponds to the
calculation of the partition function with nontrivial boundary conditions.  One can define
the generalized entropy by summing over the all boundary charges or introducing a kind of boundary
chemical potentials for different charges. Note that an example of the entanglement
entropy with nontrivial boundary structure has been discussed in \cite{sakai}. In that case
an example of a wall between two conformal theories \cite{walls} has been analyzed;
however, no charges have been attributed to the boundaries. In our case the treatment
of the generalized entropy in the conformal situation should involve dyonic
boundary conditions formulated in \cite{affleck}.

Our recipe for the holographic calculation of the generalized entanglement entropy is very
transparent. One has just to calculate of the area of the composite minimal surface as the function of the
geometrical characteristics. Since all boundaries generically have $(p,q)$ electric and magnetic charges,
the corresponding boundary contour has to be a boundary of the $(p,q)$ string worldsheet.
Such connected composite surfaces may exist or not depending on the geometry of the boundary regions.
Similar to canonical entanglement entropy,  charged entanglement entropy is UV divergent but
the UV divergent part is independent of the
geometrical factors.

A natural question concerns the properties of the generalized entropy. The first one
to be mentioned is the strong subadditivity which can be simply tested in the holographic picture.
A comparison of the corresponding area indicates that for the simplest (0,1)-(1,0)
correlator this property is satisfied, however, the analysis of the multiple $(p_i,q_i)$
loop correlators deserves a special consideration. The most interesting
question related to the generalized entanglement entropy concerns its
modular properties. Indeed, when we have a correlator of multiple dyonic
loops, it takes values in $SL(2,Z)^{\otimes k}$ with some integer $k$ and it would be very interesting
to investigate the action of the $S$-duality group on it, which could be
related to the deconfinement phase transition  \cite{klebanov}. We hope to investigate
this issue in a separate work.

\section{Conclusion}

In this paper we have considered the simplest correlator of nonlocal
electric and magnetic probes in $\mathcal{N}=4$,  that is, a Wilson and a t'Hooft loop.
The calculation has been performed at strong coupling regime and we have
identified a composite minimal surface in the curved space with the
proper boundary conditions. It turns out that the connected minimal surface
exists in an interval of the radii of the loops and there exists a kind
of phase transition similar to the one found in \cite{olesen}.
We have
also investigated the properties of the solution in the thermal background.
At low temperatures the admissible bounds for the radii of the loops
are $0.6 <R_2/R_1\leq 1$, whereas at high temperatures  above the
Hawking-Page transition $z_0 <R_2/R_1\leq 1$.
A critical 't Hooft radius always exists, which denotes a phase
transition between a disconnected configuration for radii
below it and a connected configuration above it.

We have focussed on the connected minimal surface contribution. However,
in the range of the radii when it does not exist the correlator is saturated
by the exchange of the particular supergravity mode and it would
be interesting to investigate this contribution in details as well.
It would be also interesting to recognize the phase transition
in terms of the summation of the perturbative series in the spirit of
\cite{zarembo}. However in the case under consideration the perturbation
analysis is more involved since the interactions between electric
and magnetic objects have to be summed up.

One of the most interesting questions concerns the action of the
S-duality group on the generic correlators of the dyonic $(p,q)$
loops. A generic correlator of $(p_1,q_1)$, $(p_2,q_2)$
dyonic loops has to possess interesting properties  under
the action of~$SL(2,Z)\otimes SL(2,Z)$ group. In particular,
it would be interesting to investigate the modular properties
of phase transition points of
dyonic loop correlator.

Calculation of a correlator of several nonlocal observables
has a lot in common with the calculation of the
entanglement entropy. Our calculation suggests the natural
generalization of the entanglement entropy notion to
the case when the boundaries of the regions are charged under
the  $S$-duality group. That is, generically the generalized
entanglement entropy for the region with  $k$
boundaries takes values in
the group tensor product $SL(2,Z)^{\otimes k}$. Since the entanglement entropy
at strong coupling is similar to the
Bekenstein-Hawking black hole entropy the generalized
entanglement entropy can be considered as an analogue of charged black hole entropy. We plan to discuss these issues
elsewhere.

\section*{Acknowledgements}
We are grateful to F. Gubarev and V. Zakharov for the useful
discussions. The research was supported in part by the grants
PICS-07-0292165(A.G.), RFBR 09-02-00308(A.G.), RFBR
07-01-00526(A.Z.), RFBR Grant No. 07-02-00878 (A.M.) and by the
Scientific School Grant No. NSh-3036.2008.2 (A.M.). This work was
supported by the DFG Cluster of Excellence MAP (Munich Centre of
Advanced Photonics) (A.Z.). A.G. thanks FTPI at University of
Minnesota where the part of the work was done for the kind
hospitality and support.

\end{document}